# Electron and phonon spectra dynamics and features of phase transitions in sodium at $P$ = 0 - 100 GPa

## Volkov N B[1] and Chingina E A[1,*]

[1]Institute of Electrophysics, Russian Academy of Science, Ural Branch, Amundsen Street 106, Yekaterinburg 620016, Russia

[*]zhenya@iep.uran.ru

**Abstract.** Electron and phonon spectra dynamics, as well as features of structural phase transitions and melting of sodium under pressure within range 0 to 100 GPa are investigated. Electron and phonon spectra of crystal sodium are calculated *ab initio* within the density functional theory by means of the software package LmtART-7 (see [1-3] and references herein), allowing the fully potential method of linear muffin-tin orbitals (FP-LMTO method). Earlier this method was used in papers [4-6] for calculation of metals band structure within atomic-spheres approximation (LMTO-ASA method). Using the Lindemann measure and the calculated phonon spectra the theoretical values of melting points corresponding to the experimental data are obtained. Features of electron and phonon spectra dynamics in the melting curve maximum vicinity and within structural transition range are discussed cI2 → cF4.

## 1. Introduction

Modern methods of research of structure of substances with high pressures [7, 8], allowed to find not simple behavior of simple metals (e.g. see reviews [9-12] and references therein). First, low - symmetric and incommensurate host-guest of crystal structures arose from structural phase transitions. Secondly, it is the abnormal character of curves of melting $T_m(P)$, showing existence of maxima and minima and also - sites in the curve with $dT_m/dP < 0$ [13, 14]. In the third, it is superconductivity [15, 16] and transitions metal-to-semiconductor [17] and semiconductor-to-metal [18] in Li, and transition metal-insulator in Na [19].

In experiments [13, 14] the anomalous behavior of a curve of melting $T_m(P)$ of sodium is revealed: maximum of $T_m = 1000$ K with $P \sim 30$ GPa and the subsequent its falling to room temperatures at $P = 118 \div 120$ GPa (see curves 1 and 2 designed by star symbols at Fig. 1a). The negative melting slope of sodium is in a pressure range in which the stable solid structures are known to be body-centered cubic (b.c.c. or cI2 in Pearson notation [20]) and, above 65 GPa [21], to be the more compact face-centered cubic (f.c.c. or cF4 in Pearson notation [20]). Under more pressure $P = 115$ GPa, after transition from cF4 to less symmetric cubic structure with a 16-atom cell (cI16) [13, 21], temperature of melting of $T_m = 300$ K that lower of melting at a normal pressure 371 K. See also Fig. 1a, where the sodium structure in the melting curve minimum vicinity according to crystallographic researches of authors [14] is shown on a boxed insert. It can be seen, that the transition range from crystal structures cI16 and oP8 at $P = 118$ GPa and $T = 296.6$ K to liquid with the diffuse maximum of X-ray radiation scattering equals to only $\Delta T = 9.4$ K temperature interval. Herein, within the range $296.6 \le T \le 301.4$ K the sodium structure represents a mixture of crystal structures - tetragonal tI50 and a rhombic oC120 containing 50 and 120 atoms in unit cells respectively, and within the range $301.4 \le T \le 306$ K - monoclinic structure mP512 containing 512 atoms in the cell. Therefore one may assume that the transition range, bounded from below of the melting curve passing through (306 K, 115 GPa), (296.6 K, 118 GPa), (306 K, 119.4 GPa) points, and from above with a liquid melt, represents rather polycluster amorphous body (see [24]), than mono - or a polycrystalline body. Along the line $T_m = 306$ K both density and volume of liquid are non-uniform. Thus the derivatives $\partial V_{Lm}/\partial P|_{T_m=const}, \partial V_{Sm}/\partial P|_{T_m=const}$ are negative and cross each other at $P = 118$ GPa, where $V_{Sm} = V_{Lm}$.

It should be mentioned, that since cesium melting curve anomalous behavior has been experimentally discovered in 60's of the 20[th] century [25, 26] (see also reviews [27, 28] and references therein) numerous attempts of its theoretical explanation have been performed: e.g. see [29-34]. In papers [29-31] the solid-liquid transition is considered as an order-disorder transition within Lennard-

Jones and Devonshire model of liquids [35, 36] under Bragg-Williams approximation which is, technically, applicable to alloys (solutions) only. Despite the faint applicability of Lennard-Jones and Devonshire theory and Bragg-Williams approximation for simple monoatomic liquids, such as melts of alkali metals are (see some reasoned criticism of the theory in J.I. Frenkel's book [37]), in noted papers qualitative confirmation of the chance of maximum emergence, resulting from long-range order deviation in solids, is obtained. For the cause of maximum formation on melting curve papers [32-34] calls repulsive potential and its softness to attention and also point at possible dependence of the vicinity. I. e. repulsive potential probable relation to near-order local structure of solid metal and its melt. Paper [38] should also be noted, because it says that substance model used in [29-34] cannot provide any clear statement for melting curve abnormal behavior. The authors of [38] rather convincingly showed the possibility for a model fluid having only one type of local structure to have a wide features range of melting, structural transitions in solids under condition of interparticle repulsion extreme softening, leading to nearest neighbors division reduction. In other words, extreme softening of interparticle repulsion promotes fluid clustering (polycluster amorphous body in terms of A.S. Bakai [24]). For the extra proof lets consider liquid phenomenological model suggested in our papers [22, 39]. According to the model liquid consists of the clusters having quasicrystal short-range ordering and the intercluster space, occupied with monoatomic gas (monoatomic clusters). A sodium melting curve (see curve 4 on Fig. 1a) and a curve of the pressure relative change at melting (see curve 2 on Fig.2a) were obtained within two-phase approximation. Comparison with experimental data shows, that up to 15 GPa good coincidence between theoretical and experimental melting curves is observed. Melting temperature growth deceleration and volume jump reduction acceleration are observed under pressure increasing. Though the fact is not sufficient for gaining zero-value volume jump at $P = 28.3$ GPa. See for example curve 3 on Fig. 3a (marked with star symbols), the extrapolation to zero we obtained for the experimental data represented in [28] (the curve 1 marked with triangles on Fig. 2a). In our estimation, it is concerned with both the actual two-level model and with the identical number of atoms in clusters, thus not allowing to obtain more dense, in comparison with a solid body, packing of atoms in its melt, which is observed on the sites of the melting curve with the negative slope.

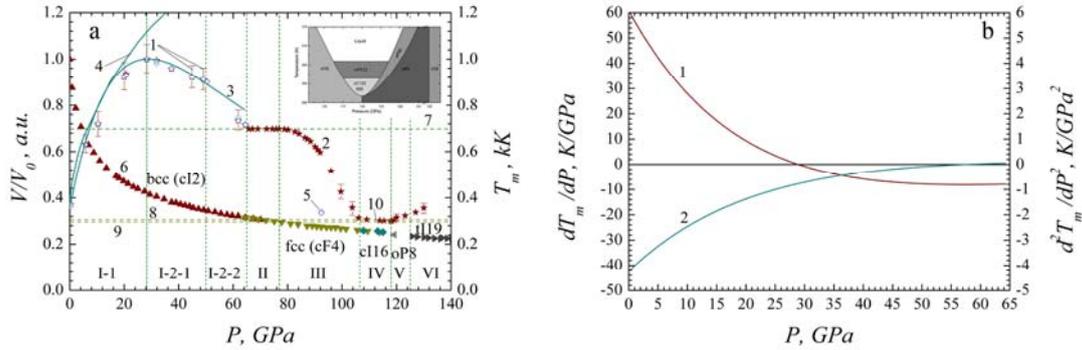

**Figure 1.** Left (a): Points (1, 2 - stars) - Na melting temperature experimental values according to [13]; curve 3 - our approximation of [13] data within the range $0 \leq P \leq 64.5$ GPa; curve 4 - melting temperature calculated values, obtained in previous paper [22] where genuine two-phase equation of state was used; points (5, diamonds) - melting temperature calculated values, obtained in current paper (onward introduced); triangles (6) represent crystalline sodium relative atomic volume and its structure variations at pressure rise [13, 14, 23]. On a boxed insert part of the melting curve minimum is shown with transition range structures being indicated according to [14]. Horizontal dash line (7) corresponds to $T_m$ within transition range cI2 → cF4, where $V_{Sm} \simeq V_{Lm}$; line (8) corresponds to $T_m$ of Na having cI16 structure (see boxed insert); line (9) corresponds to the temperature, below which sodium stays in crystalline state with structure, marked on figure with vertical dashed lines; (10) - range, corresponding to the melting curve minimum, where Na liquates from being solid. Right (b): First (1) and second (2) approximation (curve 3 on Fig. 1a) derivatives.

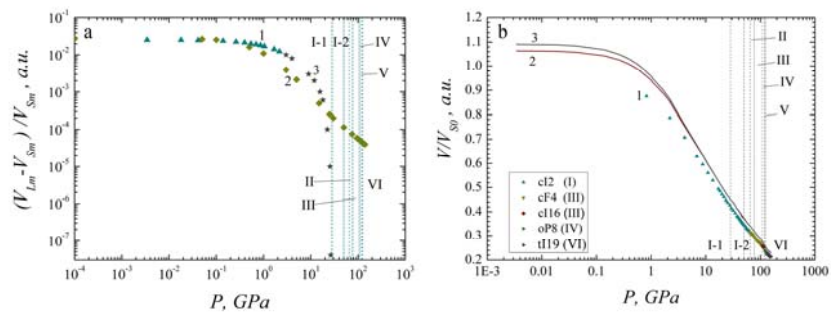

**Figure 2.** Left (a): Relative volume variation of Na while melting within the range I-1 according to experimental date from [28] - triangles (1), our calculations from [22] - diamonds (2) and our extrapolation to zero of data provided by [28] - stars (3). Right (b): Relative volume of solid and liquid sodium qualitative behavior versus pressure according to the experimental data and our interpolation (see Fig. 2 (a)).

At present time in order to describe the melting curves behavior the so called *ab initio* methods are used, in which forces acting on atoms to obtain their instantaneous arrangement (well-ordered or not) are computed within quantum dynamical methods [40-47]. For example the *ab initio* molecular dynamics simulation of Na solid and liquid phases are carried out in papers [40, 41], to find out coherence between structure and density and thus to determine the reasons of melting curve negative inclination. In papers [42, 43, 45] the explanation of sodium melting is suggested via *ab initio* calculations of phonon spectrum and mean square deviation of atoms, where only matter solid phase properties are used. The cited papers show the rebuilding of sodium being compressed phonon spectrum essential role for melting curve abnormal character formation within single-phase approximation, using well-known Lindemann measure to be melting criterion. In our opinion, not enough attention is paid to the analysis of lattice atoms zero fluctuations contribution to the features of melting curve behavior, because under pressure the lattice atoms zero fluctuations energy increases. Thus at close to zero temperatures, the equilibrium locations of atoms may not coincide with the ideal lattice nodes. In favor of this assumption it is possible to adduce cI16 equilibrium crystal structure which can be superposed of eight cI2 cells, however atoms of the latter would not reside in nodes of the former.

The analysis is also necessary because the anomalous behavior of melting curves under pressure increase may occur not only at alkali metals, but also at other metals, including noble and transitional, but under significantly greater pressure values. For instance in [48] structural transition in Au from cubic centered (fcc) to hexagonal-close-packed (hcp) structure at $P = 240$ GPa has been found experimentally; in [49] has been observed the same transition for Al at $P = 217 \pm 10$ GPa. The theoretical paper [50] suggested possibility of existence for Al at $P = 3.2 - 8.8$ TPa of an incommensurate host-guest (h-g) structure *Al16-I4/mcm* being analogous to the incommensurate host-guest structure of phase IVa of barium.

The objective of the current paper is careful analysis of electron and phonon spectra dynamics and of lattice atoms zero fluctuations contribution to cI2→cF4 structural transition features and melting curve behavior for sodium within pressures range from 0 to 100 GPa.

## 2. Calculation method

We used the LMTART-7 software package [2, 3] for *ab initio* phonon and electron spectra determination, as well as of a total interaction energy between electrons and sodium nuclei. Dynamics of a lattice was computed using linear response method within well-known density functional theory (see details of the method in reviews [51-53] by LMTART-7. The exchange-correlation potential (Vosko-Wilk-Nussair) [54] was considered in a generalized gradient approximation [55]. We used a pseudo-potential method in plane wave expansion (PLW). The PLW expansion is a full potential approximation which uses non-overlapping muffin-tin spheres, where the potential is represented via spherical harmonics, expansion, and the interstitial region where the potential is expanded in plane waves. The full potential regime provides the best accuracy at the price of increasing computational time [2, 3]. The number of plane waves was used 14146, the plane-wave energy cutoff was 138.1 Ry (up to 250 Ry). Integration on Brillouin zone was carried out on a grid 32*32*32 special (Monkhorst-Pack) k-points. Phonon spectra were calculated by means of the interatomic power constants on the basis of 35 basic q-vectors in the irreducible Brillouin zone (a grid 8*8*8). The calculated via LMTART-7 cI2 Na phonon states densities for various pressure values correspond to normalization requirements within the accuracy substantially smaller than one percent. The highest deviation, 1.6%, is obtained for $P = 0.263$ GPa.

To obtain Na melting temperature, along with authors of [42, 43, 45], we use well-known Lindemann measure

$$\sqrt{\ll u^2(T_m) \gg} = L d_{NN},$$ (1)

where $\ll u^2(T_m) \gg$ - atoms displacement standard deviation at melting temperature $T_m$; $L, d_{NN}$ - Lindemann measure and mean nearest-neighbor distance correspondingly.

Atoms displacement standard deviation depends on normalized phonon states density $g_{ph}(\omega)$ via relation [56]:

$$\ll u^2(T) \gg = \frac{\hbar}{2M} \int \frac{g_{ph}(\omega)}{\omega} \coth \frac{\hbar \omega}{2T} d\omega.$$ (2)

Normalization requirement $g_{ph}(\omega)$ is:

$$\int_{\omega_{min}}^{\omega_{max}} g_{ph}(\omega) d\omega = 1,$$ (3)

where $g_{ph}(\omega_{min}) = g_{ph}(\omega_{max}) \equiv 0$. It should be noted that $\omega_{min}$ values, obtained via LMTART-7 software package, might become negative (results of negative frequencies minimal values emergencies will be further analyzed). Herein, according to our calculations, the normalization requirement (3) is fulfilled highly accurate.

At $T \gg \hbar \omega$ equation (2) transforms to:

$$\ll u^2(T) \gg = \frac{T}{M} \int \frac{g_{ph}(\omega)}{\omega^2} d\omega.$$ (4)

The resultant expressions we used for $T_m$ calculations in both quantum case and classical limit are written as:

$$(L d_{NN})^2 - \frac{\hbar}{2M} \int_{\omega_{cut}}^{\omega_{max}} \frac{g_{ph}(\omega)}{\omega} \coth \frac{\hbar \omega}{2T} d\omega = 0;$$ (5)

$$T_m = (L d_{NN})^2 M \left( \int_{\omega_{cut}}^{\omega_{max}} \frac{g_{ph}(\omega)}{\omega^2} d\omega \right)^{-1}.$$ (6)

The replacement of $\omega_{min}$ in (5) and (6) with $\omega_{cut} > \omega_{min}$ is caused by infeasibility of physically sensible data to be obtained while curve $T_m(P)$ calculations within full frequency range $[\omega_{min}, \omega_{max}]$ ($\omega_{cut}(P), \omega_{max}(P)$ behavior is to be analyzed further). We should note that our $\omega_{max}$ values calculated via LMTART-7 are close to the ones obtained in [43] via Quantum ESPRESSO software package [57].

Calculations of $T_m$ using (5) and (6) were performed in MATHEMATICA 7 software package. For Na taken Lindemann constant value was $L$=0.14 similar to [43].

## 3. Results. Discussion

Important characteristic of FP-LMTO method is the non-touching muffin-tin sphere radius, $R_{MT}$, it separates electrons belonging to an atom core from electrons of external orbitals. Fig. 3 shows the pressure dependent relation $R_{MT} / R_c$ ($R_c$ - average nearest neighbors distance determining the macroscopic density (i.e. volume)). It can be seen that for the given lattice type the value is constant (curve 2), while the relative volume per atom (curve 1) smoothly decreases. Within the region II, where structural transition cI2→cF4 takes place, it decreases abruptly. Inside this region, at least within the range $P \approx 65 - 75$ GPa, the phases cI2 and cF4 simultaneous existence is observed.

Let's return to Fig.1. We obtained the analytical approximation of the experimental results in region I (cI2 lattice, curve 3 on Fig.1a) as follows:

$$T_m = T_{m0} + A \left( \exp \left( -\alpha_1 \left( P - P_0 \right) \right) - \exp \left( -\alpha_2 \left( P - P_0 \right) \right) \right),$$ (7)

where $T_{m0} = 371$ K; $A = 1.70981 \cdot 10^{8}$ K, $\alpha_1 = 3.49197 \cdot 10^{-2}$ GPa$^{-1}$; $\alpha_2 = 3.492 \cdot 10^{-2}$ GPa$^{-1}$. The first and the second derivatives of the melting curve are shown on Fig.1b. It is obvious that the $T_m$ maximum point dos not belong to simple second-order phase transition, if existence of the implicit local order

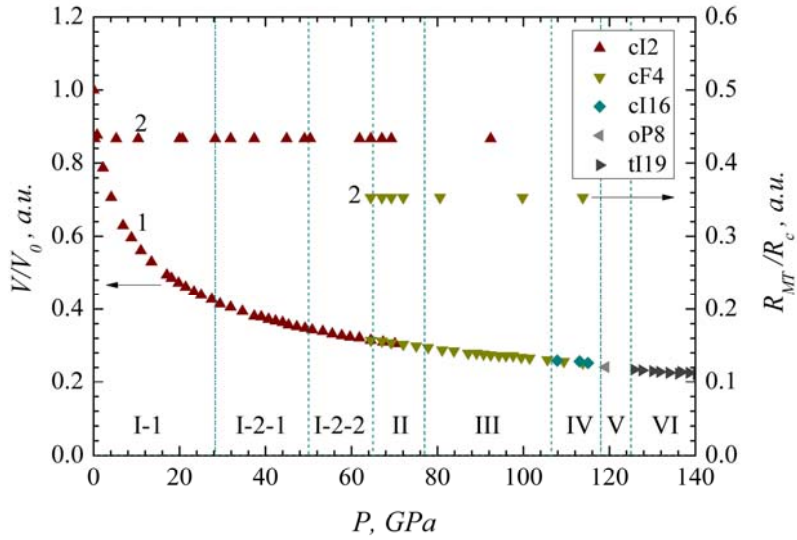

**Figure 3.** Relative volume [23] (1) and $R_{MT}/R_c$ ratio calculated via LmtART-7 [2] dependences versus pressure comparison.

parameter is not assumed, like suggested as tensor irreducible part in paper [58]

$$T_{\alpha\beta\gamma\delta}(\mathbf{x}) = \sum_{i=1}^{3} n_{\alpha}^{i}(\mathbf{x})n_{\beta}^{i}(\mathbf{x})n_{\gamma}^{i}(\mathbf{x})n_{\delta}^{i}(\mathbf{x}), \qquad (8)$$

where orthogonal unit vectors $\{\mathbf{n}^i\}(\mathbf{x})$ $(i = 1, 2, 3)$ define the local cubic order direction in point $\mathbf{x}$. The real situation is apparently much more difficult. First, because the order parameter (8) can be applied to a simple cubic lattice only. Secondly, the melting curve and a dashed vertical line at $P = 28.3$ GPa divide the phase plane into four areas with various local densities and a symmetries. Within the region I-1 metal density below $T_m(P)$ is greater, than the melt density above $T_m(P)$. Within the region I-2 - situation is vice versa: above $T_m(P)$ the melt density exceeds metal density below $T_m(P)$. Unfortunately, the single-phase Lindemann measure along with *ab initio* molecular dynamics methods are not capable of solving the problem uniquely. The calculated values (diamonds, Fig. 1a) we obtained over relations (5) and (6) are well consistent with experimental data. However this agreement is achieved by cutting off the initial frequencies of phonon states density (see curve 2 on Fig. 4a). Moreover, the behavior of curve $\omega_{cut}(P)$ qualitatively repeats the melting curve behavior, with just more flat maximum being shifted to the side of larger pressures. It indicates that the melting curves data is to be accepted cautiously when obtained via Lindemann measure, or by similar molecular dynamics methods. In our opinion, while melting curves at high pressures abnormal behavior investigation, it is necessary to pay more attention to the research on condensed matter structure and its near order symmetry.

Fig. 4b shows the dependence of the zero fluctuations energy relation to the latent melting heat at normal conditions, obtained within the Debye model of metal with boundary frequency $\omega_{max}(P)$ (curve 1) and

$$\varepsilon_{vac} = \frac{\hbar}{2} \int_{\omega_{min}}^{\omega_{max}} \omega g_{ph}(\omega)d\omega$$

(curve 2). It is obvious that in region I the vacuum fluctuations energy can reach up to 20% of latent melting heat thus encourage disorder growth when with pressure rises. The suggested conclusion assigns necessity of considering zero fluctuations contribution to both cases of lattice long range order destruction while melting, and also of crystalline structure reconstruction at high pressures. The additional pro argument can be found in paper [59], it claims that growth of pressure promotes lattice fluctuations anharmonicity increase, which suppresses superconductivity and leads to superconductor-to-semiconductor transition in lithium at $P = 80$ GPa (See [18] and references therein).

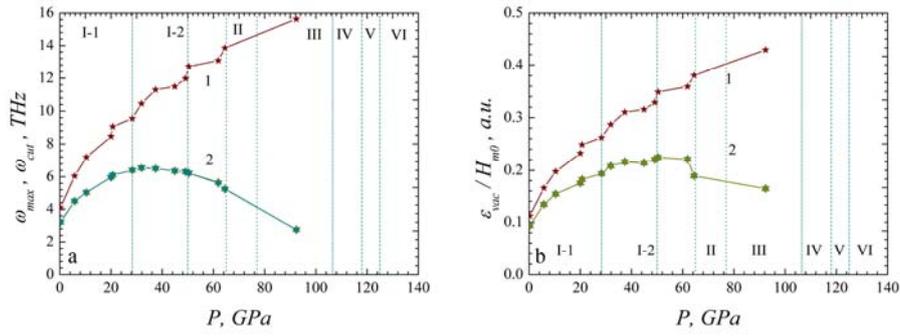

**Figure 4.** Values of boundary (1) and cutoff (2) frequencies of phonons (a) and vacuum fluctuations energy to latent melting heat ratio at normal conditions(b). Curve 1 is calculated within Debye model approximation taking into account the boundary frequency value; curve 2 is obtained according to phonon states density computed via LMTART-7 [2].

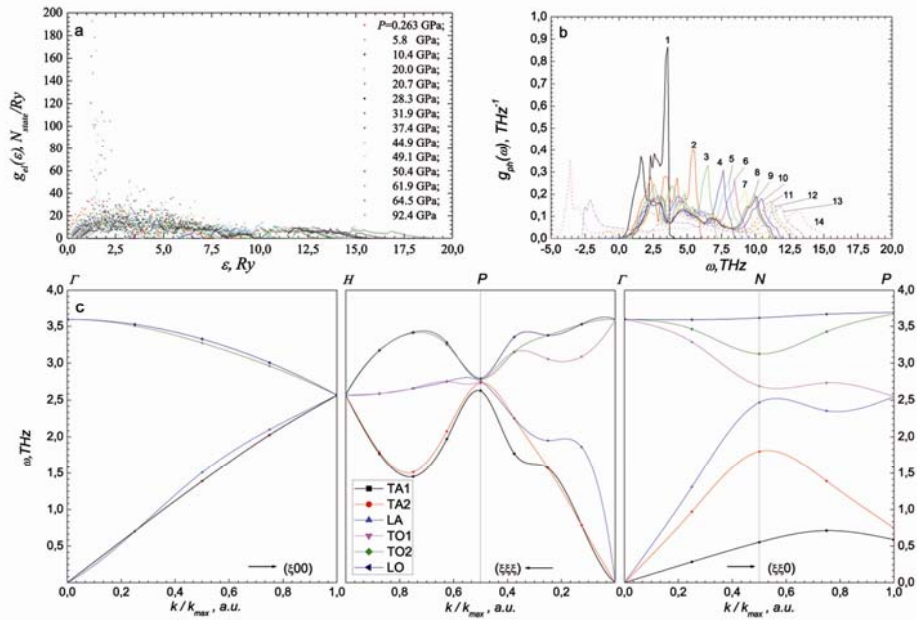

**Figure 5.** Electron (a) and phonon (b) states density computed via LMTART-7 [2], and also phonon spectrum (c) for $P = 0.263$ GPa.

Let's consider dynamics of phonon spectrum and density of electron and phonon states near the melting curve maximum and at the beginning of phase transition cI2→cF4. Fig. 5 shows electron (a) and phonon (b) states densities, and also phonon spectrum for $P = 0.263$ GPa (c). The curves numbers on Fig. 5b correspond to pressure values designed on Fig. 5a. As all calculations were made for the unit cell containing two atoms, three acoustic branches and also three optical branches do exist. Our calculated acoustic spectrum qualitatively coincides with the sodium spectrum given in textbooks, e.g. [60].

Fig. 6 shows electron and phonon densities of states (a) and a phonon spectra (b, c, d) in the vicinity of $T_m$ maximum (left); and electron (for b.c.c. and f.c.c. lattices) and phonon densities of states for b.c.c. lattice at $P = 64.494$ GPa (a), phonon spectra of b.c.c. lattice for three pressure values: (b) 61.9 GPa, (c) 64.494 GPa and (d) 92.371 Gpa (right). The features of phonon spectra in the vicinity of melting curve maximum are non-zero positive frequency value existence at $k = 0$ and small negative transverse acoustic branch frequency at $k/k_{max} \simeq 0.2$. It demonstrates the sodium local structure transformation. The existence of larger segments at acoustic and optical branches lying in frequencies negative values region testifies the instability of these branches and of lattice itself.

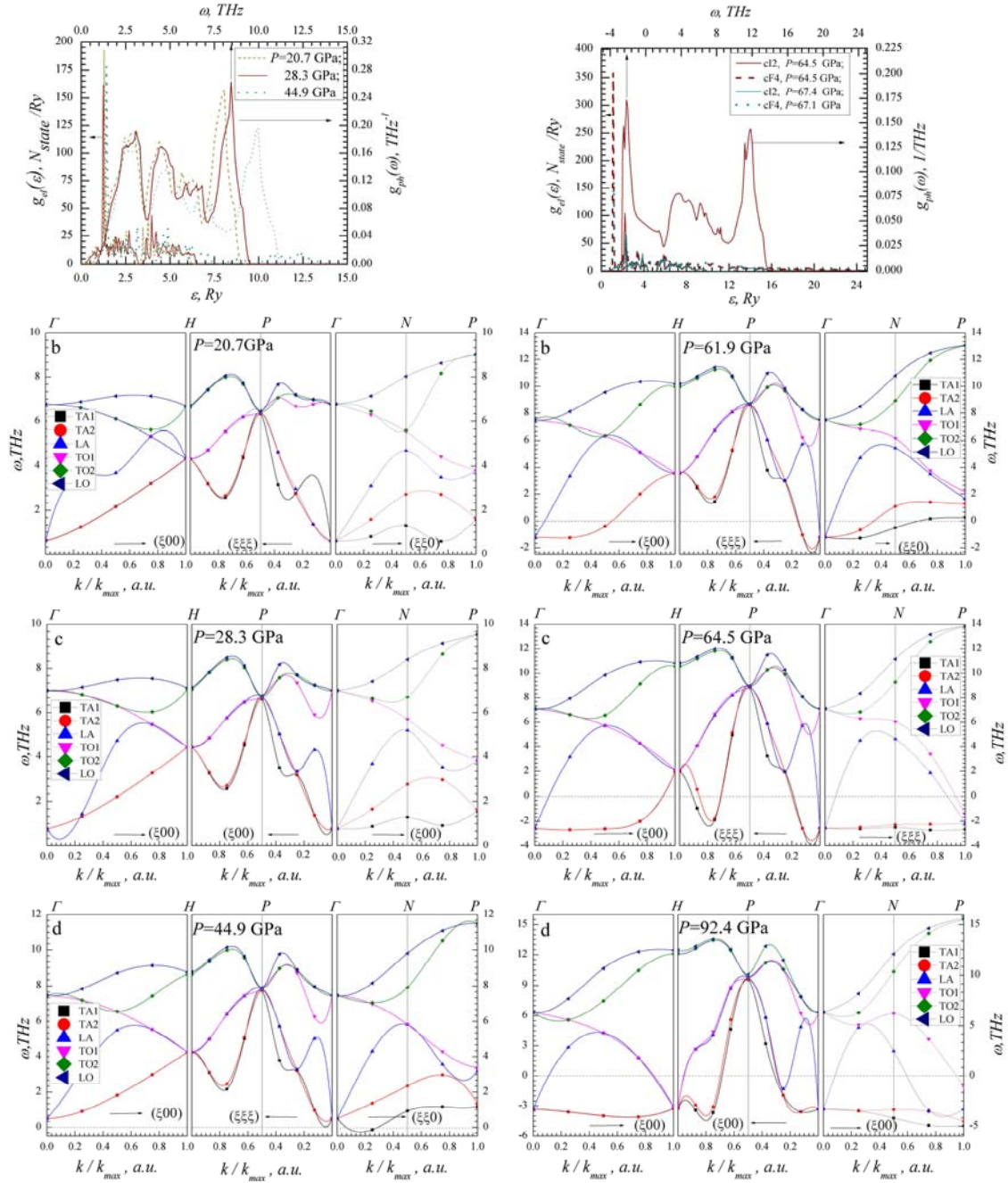

**Figure 6.** Left: Electron and phonon densities of states (a) and a phonon spectra of Na b.c.c. lattice in the vicinity of melting curve maximum:(b, c, d); Right: Density of electron b.c.c. and f.c.c. lattices states and phonon b.c.c. lattices states for sodium in the range of b.c.c.-f.c.c. transition (a), phonon spectrum of sodium b.c.c. lattice in the vicinity of transition b.c.c.-f.c.c. (b, c) and in the range of only f.c.c. lattice existence (see also Fig. 1).

## 4. Conclusion

Thus in present paper we obtained energy distribution of electrons and phonons and density of states for all experimental points within the range of Na b.c.c. lattice existence. Using the one-phase Lindemann measure and the calculated phonon spectra the theoretical values of melting points matching the experimental data are obtained. Energy of zero fluctuations is defined and the necessity of its contribution to the dynamics of electron and phonon spectra to be accounted is shown. Features of volume jump behavior during melting under pressure increase are discussed.

## Acknowledgments

The authors appreciate S.Y. Savrasov for providing the LMTArt-7 software package for electron and phonon spectra of sodium calculations at various pressures, as well as K.A. Nagayev for help in preparing of English version of the article. The work is carried out within the state order No. 0389-2014-0006 and under the partial financial support of the RFBR (project No. 16-08-00466) and the Ural Branch of RAS within the UB RAS fundamental research program "Matter at high energy densities" (project No. 15-1-2-8).